\documentclass[prb,twocolumn,showpacs,amsmath,amssymb]{revtex4}
\usepackage{graphicx,ulem}
\usepackage{enumerate}
\newcommand \beq{\begin{eqnarray}}
\newcommand \eeq{\end{eqnarray}}

\newcommand{\bfr}{\mathbf{r}}
\newcommand{\bfR}{\mathbf{R}}
\newcommand{\bfk}{\mathbf{k}}
\newcommand{\bfK}{\mathbf{K}}
\newcommand{\bfG}{\mathbf{G}}
\newcommand{\bfs}{\mathbf{s}}

\newcommand{\atau}{a_{\tau'}}
\newcommand{\vf}{v_{_F}}

\newcommand{\Tr}{\mathrm{Tr}}

\begin{document}

\title{Phase structure of monolayer graphene from effective U(1) gauge theory on honeycomb lattice}
\author{Yasufumi Araki}
\affiliation{
Department of Physics, The University of Tokyo, Tokyo 113-0033, Japan
}

\begin{abstract}
Phase structure of monolayer graphene is studied on the basis of 
 a U(1) gauge theory defined on the honeycomb lattice.
Motivated by the strong coupling expansion of U(1) lattice gauge theory,
we consider on-site and nearest-neighbor interactions between the fermions.
When the on-site interaction is dominant,
the sublattice symmetry breaking (SLSB) of the honeycomb lattice takes place.
On the other hand,
when the interaction between nearest neighboring sites is relatively strong,
there appears two different types of spontaneous Kekul\'e distortion (KD1 and KD2),
without breaking the sublattice symmetry.
The phase diagram and phase boundaries separating SLSB, KD1 and KD2 are obtained from the 
 mean-field free energy of the effective fermion model.
A finite gap in the spectrum of the electrons can be induced in any of the three phases.
\end{abstract}
\maketitle

\section{Introduction}
The discovery of graphene \cite{Novoselov_2004},
a one-atom thick material of carbon atoms,
has made a great impact
not only on condensed matter physics but also on particle physics \cite{CastroNeto_2009}.
It gives a realization of massless quasiparticles in a material easy to create and observe,
since the valence band and the conduction band of the electrons
touch at two independent ``Dirac points'' in the Brillouin zone
with the conical shape \cite{Wallace_1947}.
Due to this ``Dirac cone'' structure,
there can be seen several unconventional behaviors characteristic to monolayer graphene,
such as the high mobility of charge carriers and the half-integer quantum Hall effect.
Since these charged quasiparticles obey the Dirac equation around half filling,
they are described as massless Dirac fermions in the (2+1)-dimensional plane,
as an effective field theory \cite{Semenoff_1984}.
Such an effective field description also has some connections to the high energy physics side,
such as lattice fermion formulation \cite{Creutz_2008}, deformation-induced gauge fields \cite{Jackiw_2007,Sasaki_2008,Vozmediano_2010} and the existence of vortex zero modes \cite{Hou_2007,Seradjeh_2008}.

The effect of the Coulomb interaction between electrons is one of the most important problems
in graphene physics \cite{Physics_2009}.
Since the Coulomb interaction strength in graphene is effectively enhanced
by the inverse of the Fermi velocity $\vf \sim c/300$
from the ordinary quantum electrodynamics (QED),
it is beyond the treatment of perturbative expansion
unless the interaction is screened by dielectric substrates such as silicon oxides ($\mathrm{SiO_2}$).
If the interaction is sufficiently strong,
the electron and hole may form an exciton condensate,
which may give a finite gap in the band structure of graphene.
This scenario is analogous to the dynamical mass generation of fermions
in strongly coupled gauge theories such as quantum chromodynamics (QCD),
where the spontaneous breaking of the chiral symmetry
leads to the dynamical mass gap of the fermions \cite{Hatsuda_Kunihiro_1994}.
In the effective field theory of graphene,
the chiral symmetry of the fermions corresponds to
the inversion symmetry between two triangular sublattices of the honeycomb lattice.
Owing to such an analogy,
there have been several studies on the ``chiral symmetry breaking'' in monolayer graphene
with the techniques commonly used in the studies on QCD.
Schwinger--Dyson equation \cite{Gorbar_2001,Gorbar_2002,Khveshchenko_2001}, $1/N$ expansion \cite{Son_2007,Herbut_2006},
and the exact renormalization group approach \cite{Giuliani_2010} have been applied
to the effective field theory of monolayer graphene.
Monte Carlo simulations of the effective square lattice model
have been performed to obtain the critical value of the coupling constant
and the equation of state around the critical point \cite{Drut_2009,Hands_2008}.
The author has treated the system as a strongly coupled U(1) lattice gauge theory
by the strong coupling expansion,
which is one of the methods to investigate the non-perturbative features of the strongly coupled gauge theories like QCD \cite{Kawamoto_1981,Drouffe_1983,Nakano_2010},
and has obtained the behavior of the (pseudo-)Nambu--Goldstone mode
related to the chiral symmetry breaking in the low energy region \cite{Araki_2010}.

In graphene, however,
there may be other ordering patterns than the sublattice (chiral) symmetry breaking
that may open a finite spectral gap,
due to the honeycomb lattice structure \cite{Nomura_2009,Raghu_2008}.
Kekul\'e distortion,
which is described by the alternating pattern of the bond strengths
like in the benzene molecule \cite{Viet_1994},
is one of those ordering patterns without breaking the sublattice symmetry.
It can be induced externally by the effect of some substrates \cite{Farjam_2009}
or adatoms on the layer \cite{Cheianov_2009}.
In the author's previous work,
it has been found that sufficiently large external Kekul\'e distortion
may restore the sublattice symmetry
which has been spontaneously broken in the strong coupling limit of the Coulomb interaction \cite{Araki_2011}.
On the other hand,
there has been an argument that the Kekul\'e distortion may appear spontaneously
as a result of the electron-electron interaction \cite{Hou_2007,Weeks_2010}.
It is currently a great challenge what order may appear in the vacuum-suspended graphene
due to the effectively strong Coulomb interaction.
In order to treat the ordering patterns
characteristic to the honeycomb lattice exactly,
the analysis of the effective field theory preserving the honeycomb structure
is required \cite{Chakrabarti_2009,Giuliani_2012,Brower_2011}.

In this paper,
we investigate the competition between two phases,
the sublattice symmetry broken (SLSB) phase and the Kekul\'e distortion (KD) phase,
by using an effective fermion model of graphene keeping the original honeycomb lattice structure.
This model, motivated by the strong coupling expansion of U(1) lattice  gauge theory 
 defined on the honeycomb lattice,
includes the on-site interaction and the nearest-neighbor (NN) interaction.  
 If the on-site interaction is dominant,
the SLSB of the honeycomb lattice takes place.
On the other hand, if the interaction between nearest neighboring sites is relatively strong,
there appears two types of spontaneous Kekul\'e distortion (KD1 and KD2),
without breaking the sublattice symmetry.
By analyzing the mean-field free energy of the the effective fermion model, 
 we derive the phase diagram and phase boundary separating the three phases,  SLSB, KD1 and KD2.
 A finite gap in the spectrum of the electrons can be induced in  any of the three phases.

This paper is organized as follows. 
In Section \ref{sec:model},
we construct U(1) lattice  gauge theory 
on the honeycomb lattice starting from the conventional tight binding Hamiltonian
coupled with the electromagnetic field as compact U(1) link variables.
In Section \ref{sec:expansion},
we derive the interaction terms between fermions
by using the strong coupling expansion techniques of the U(1) lattice gauge theory
up to the next-to leading order \cite{Kawamoto_1981,Drouffe_1983}.
Two characteristic interactions are induced;
the on-site interaction which favors SLSB,
and the NN interaction which favors KD.
We take an effective fermion model including these two interaction terms.
In the next two sections,
we take the effective fermion model as it is
and investigate the phase structure by varying the strength of the on-site and NN interactions,
to study the interplay between these different orders.
In Section \ref{sec:analysis},
we investigate the phase structure qualitatively
by taking two characteristic cases; on-site dominance and NN dominance.
Phase diagram with SLSB, KD1 and KD2 phases are also drawn qualitatively.  
In Section \ref{sec:numerical},
we confirm the the phase diagram in the previous section numerically by 
minimizing the mean-field free energy of the effective fermion model.  
Section \ref{sec:conclusion} is devoted to summary and concluding remarks.

\section{Gauged honeycomb lattice model}\label{sec:model}
\begin{figure}[tb]
\begin{center}
\includegraphics[width=5.5cm]{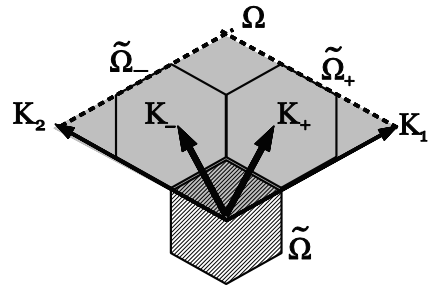}
\end{center}
\vspace{-0.5cm}
\caption{The schematic picture of the Brillouin zone $\Omega$ and the Dirac points $\bfK_\pm$.
When the Kekul\'e distortion pattern is induced,
the unit lattice in the real space is three times enlarged,
so that the Brillouin zone is split into three parts, $\tilde{\Omega}$ and $\tilde{\Omega}_\pm$.}
\label{fig:brillouinzone2}
\end{figure}

In order to construct the model action of the system preserving the original honeycomb lattice structure,
we start from the conventional tight-binding Hamiltonian \cite{Wallace_1947},
\begin{equation}
H = -h \sum_{\bfr \in A} \sum_{i=1,2,3} \left[ a^\dag(\bfr) b(\bfr+\bfs_i) + \mathrm{H.c.} \right], \label{eq:tight-binding}
\end{equation}
which describes the hopping of an electron
between nearest neighboring sites
with amplitude $h=2.7\mathrm{eV}$.
Here $a(a^\dag)$ and $b(b^\dag)$ are the annihilation (creation) operators of electrons
on the lattice sites in A and B sublattices respectively.
$\bfs_1=(0,-a), \; \bfs_2=\left(\frac{\sqrt{3}a}{2},\frac{a}{2}\right), \; \bfs_3=\left(-\frac{\sqrt{3}a}{2},\frac{a}{2}\right)$ are the hopping directions, with the lattice spacing $a=|\bfs_i|=1.42$\AA.
The triangular sublattices A and B are spanned by two lattice vectors
$\bfR_1=\bfs_2-\bfs_1$ and $\bfR_2=\bfs_3-\bfs_1$.
In the momentum space,
the Brillouin zone is spanned by reciprocal vectors $\bfK_{1,2}$,
where $\bfK_i \cdot \bfR_j = 2\pi\delta_{ij}$.
By diagonalizing this Hamiltonian in the momentum space,
the dispersion relation reveals the ``Dirac cone'' structure
\begin{equation}
E(\bfK_\pm + \bfk)=h|\Phi(\bfK_\pm + \bfk)|=v_{_F}|\bfk| + O(k^2)
\end{equation}
around two independent Dirac points $\bfK_\pm$,
where $\Phi(\bfk) \equiv \sum_{i=1,2,3} e^{i\bfk\cdot\bfs_i}$ (see Fig.\ref{fig:brillouinzone2}).
When the system is half-filled,
the valence band and the conduction band touches only at these points.
The Fermi velocity $v_{_F}=(3/2)ah=3.02\times 10^{-3}$ is considerably smaller than the speed of light.
This Hamiltonian possesses an inversion symmetry between two sublattices A and B,
which can be extended to the continuous $\mathrm{U(1)_A}$ symmetry in the low-energy region.
On-site energy difference between two sublattices,
$m(a^\dag a-b^\dag b)$, breaks this sublattice symmetry,
which corresponds to the mass term $m\bar{\psi}\psi$ of the Dirac fermions.

From the Hamiltonian in Eq.(\ref{eq:tight-binding}),
the effective action for fermions $S_F$ is derived with the imaginary time ($\tau$) formulation.
Here we perform the temporal scale transformation $\tau \rightarrow \tau'/v_{_F}$,
so that the Fermi velocity shall be rescaled to unity.
The temporal direction is discretized with the lattice spacing $\atau(=v_{_F} a_\tau)$
equal to the spatial lattice spacing $a$.
As a consequence of this discretization,
we have a pair of fermion doublers in the temporal direction \cite{Nielsen_Ninomiya_1981},
which we consider here as the spin (up/down) degrees of freedom.

In this lattice model,
the effect of the electromagnetic field is implemented by U(1) link variables
between spatially or temporally neighboring sites:
\begin{eqnarray}
S_F &=& \frac{1}{2} \sum_{\mathbf{r} \in A;\tau'} \left[a^\dag(x)U_{\tau'}(x)a(x+a\hat{\tau'})-\mathrm{H.c.}\right] \\
 && +\frac{1}{2} \sum_{\mathbf{r} \in B;\tau'} \left[b^\dag(x)U_{\tau'}(x)b(x+a\hat{\tau'})-\mathrm{H.c.}\right] \nonumber \\
 && +\frac{a h}{v_{_F}} \sum_{\mathbf{r} \in A,\tau'} \sum_{i=1}^{3} \left[a^\dag(x) U_i(x) b(x+\mathbf{s}_i) + \mathrm{H.c.} \right], \nonumber
\end{eqnarray}
where the lattice site $x \equiv (\bfr,\tau')$
and the link variables
\begin{eqnarray}
U_{\tau'}(\mathbf{r},\tau') &\equiv& \exp\left[ie\int_{\tau'}^{\tau'+a} d\tau' A_{\tau'} \right], \; \text{(temporal)}\\
U_i(\mathbf{r},\tau') &\equiv& \exp\left[ie\int_{\bfr}^{\bfr+\bfs_i} d\bfr' \cdot \mathbf{A} \right], \; \text{(in-plane)}\\
U_z(\mathbf{r},\tau') &\equiv& \exp\left[ie\int_{\bfr}^{\bfr+a\hat{z}} dz A_z \right]. \; \text{(out-of-plane)}
\end{eqnarray}
This lattice construction is similar to that employed in Ref.\onlinecite{Brower_2011}, 
while they differ in the treatment of the U(1) gauge field and the imaginary time discretization.

Using these U(1) link variables,
the kinetic term of the gauge field
\begin{equation}
S_G = \frac{1}{4} \int d^4 x \sum_{\mu,\nu=0}^{3}(\partial_\mu A_\nu - \partial_\nu A_\mu)^2
\end{equation}
can be rewritten on the honeycomb lattice as
\begin{eqnarray}
&& \!\! \!\! \!\! S_G = -\frac{2}{3\sqrt{3}g^2\vf} \sum_{\mathbf{r} \in A;\tau'} \mathrm{Re} U_\mathrm{hex} - \frac{\sqrt{3}}{g^2\vf}\sum_{\mathbf{r} \in A;\tau'} \sum_{i=1}^{3}\mathrm{Re}U_{iz} \nonumber \\
&& -\frac{\sqrt{3}\vf}{g^2}\sum_{\mathbf{r} \in A;\tau'}\sum_{i=1}^{3}\mathrm{Re}U_{i\tau'} - \frac{3\sqrt{3}\vf}{4g^2}\sum_{\mathbf{r} \in A \cup B;\tau'} \!\!\!\! \mathrm{Re}U_{z \tau'}, \label{eq:s-gauge}
\end{eqnarray}
where the QED coupling constant $g^2 = e^2/\epsilon_0 = 4\pi\alpha_\mathrm{QED}$,
and the constant terms are neglected.
The sum is taken over the (3+1)-dimensional space.
Here, the plaquette on the $(x,y)$-plane, $U_\mathrm{hex}$, is hexagonal-shaped,
while the other are square-shaped.
The plaquettes are defined as
\begin{eqnarray}
U_\mathrm{hex}(x) &\equiv& U_1(x) U_3^*(x+\bfs_1-\bfs_3) U_2(x+\bfs_1-\bfs_3) \\
 && \quad \times U_1^*(x+\bfs_2-\bfs_3) U_2(x+\bfs_2-\bfs_3) U_2^*(x) \nonumber \\
U_{iz}(x) &\equiv& U_i(x) U_z(x+\bfs_i) U_i^*(x+a\hat{z}) U_z^*(x) \\
U_{i\tau'}(x) &\equiv& U_i(x) U_{\tau'}(x+\bfs_i) U_i^*(x+a\hat{\tau'}) U_{\tau'}^*(x) \\
U_{z\tau'}(x) &\equiv& U_z(x) U_{\tau'}(x+a\hat{z}) U_z^*(x+a\hat{\tau'}) U_{\tau'}^*(x).
\end{eqnarray}
As a consequence of the temporal scale transformation,
the spatial part of the gauge field [the first line in Eq.(\ref{eq:s-gauge})]
becomes weakly coupled with the effective coupling strength $g^2\vf$,
while the temporal part [the second line] becomes strongly coupled with the strength $g^2/\vf$.
Since the coefficient of the spatial part $1/g^2\vf$ is sufficiently large,
we can apply a saddle point approximation to these two terms,
yielding a saddle point solution $U_\mathrm{hex}=U_{iz}=1$.
In other words,
the retardation effect (magnetic field) can be neglected
due to the discrepancy between the speed of light and the speed of fermions ($\vf$),
which is referred to as ``instantaneous approximation.''
We can safely set the spatial link variables $U_i$ and $U_z$ to unity by the gauge transformation,
leaving only the temporal link variable $U_{\tau'}$.
The fluctuation of the spatial link variables around the saddle point,
which can be considered by weak coupling expansion,
is not taken into account in this work.
As a result,
$S_G$ can be simplified as
\begin{eqnarray}
S_G &=& -\sqrt{3}\beta \sum_{\mathbf{r} \in A;\tau'}\sum_{i=1}^{3}\mathrm{Re} U_{\tau'}(x+\bfs_i) U_{\tau'}^*(x) \nonumber \\
 && \quad - 3\sqrt{3}\beta \sum_{\mathbf{r} \in A \cup B;\tau'} \!\!\!\! \mathrm{Re} U_{\tau'}(x+a\hat{z}) U_{\tau'}^*(x).
\end{eqnarray}
The parameter $\beta\equiv v_{_F}/g^2$ represents the inverse of the coupling strength,
which is $0.037$ in the vacuum-suspended graphene.
Here, we fix the Fermi velocity $\vf$ to the physical value
observed in the system with $\mathrm{SiO_2}$ substrate.

\section{Strong coupling expansion}\label{sec:expansion}
Here
we apply the techniques of the strong coupling expansion
to the effective action defined in the previous section,
and derive the effective interaction terms between fermions
by integrating out the gauge degrees of freedom,
to construct the effective model which may describe the interplay
between the sublattice symmetry breaking and the Kekul\'e distortion.
With the effective action $S=S_F+S_G$,
the partition function of the system is given by path integral
\begin{equation}
Z = \int [d\chi^\dag d\chi][d U_{\tau'}] \exp\left\{-S_F[\chi^\dag,\chi;U_{\tau'}] - S_G[U_{\tau'}]\right\},
\end{equation}
where $\chi=a,b$.
Since the gauge term $S_G$ is proportional to the small parameter $\beta$,
we can expand this equation around $\beta=0$ (strong coupling limit) and perform the path integral order by order:
\begin{eqnarray}
Z &=& \sum_{n=0}^{\infty} Z^{(n)}, \\
Z^{(n)} &=& \int [d\chi^\dag d\chi][d U_{\tau'}] e^{-S_F} \frac{(-S_G)^n}{n!}.
\end{eqnarray}
Here we take the terms up to $O(\beta^1)$.
Since the integrand can be written as a polynomial of $U_{\tau'}$ and $U_{\tau'}^*$,
integration by the link variables can be performed analytically.
As a result of the link integration,
two kinds of interaction terms are derived:
the on-site interaction in the leading order [$O(\beta^0)$],
and the nearest neighbor interaction in the next-to leading order [$O(\beta^1)$].
In order to convert these four-Fermi terms into fermion bilinear,
we apply the Stratonovich--Hubbard transformation
by introducing two kinds of bosonic auxiliary fields,
corresponding to the amplitude of sublattice symmetry breaking
and the spontaneous Kekul\'e distortion respectively.
By integrating out the fermionic degrees of freedom,
we derive the effective potential of the system
as a function of these order parameters.

\begin{figure}[tb]
\begin{center}
\includegraphics[width=7.5cm]{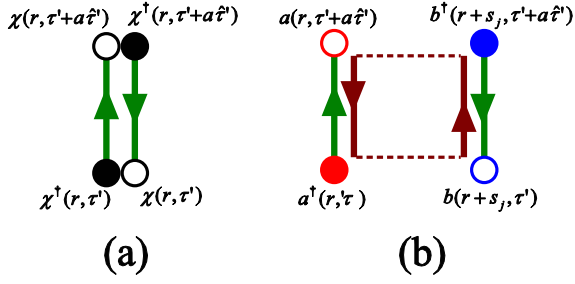}
\end{center}
\vspace{-0.5cm}
\caption{Schematic pictures of the link integration in (a) the leading order (LO) and (b) the next-to LO (NLO) in the strong coupling expansion. $(\chi=a,b)$}
\label{fig:link-integration}
\end{figure}

\subsection{Leading order: $O(\beta^0)$}
In the leading order (LO), the gauge term $S_G$ does not contribute,
\begin{equation}
Z^{(0)} = \int [d\chi^\dag d\chi][d U_{\tau'}] e^{-S_F},
\end{equation}
so that the link variables come only from the temporal hopping terms of the fermions.
On each lattice site $x=(\bfr,\tau')$,
the contribution to the link integration is
\begin{eqnarray}
&& \int dU_{\tau'} \exp \left[-\frac{1}{2}\left(\chi^\dag U_{\tau'} \chi' - \chi'^\dag U_{\tau'}^* \chi \right)\right] \nonumber \\
&=& \int dU_{\tau'} \left[1-\frac{1}{2}\chi^\dag U_{\tau'} \chi' \right] \left[1+\frac{1}{2}\chi'^\dag U_{\tau'}^* \chi \right] \\
&=& 1 + \frac{1}{4} \chi^\dag \chi \chi'^\dag \chi' = \exp\left[\frac{1}{4} \chi^\dag \chi \chi'^\dag \chi'\right],
\end{eqnarray}
where $\chi'=\chi(x+a\hat{\tau'})$.
The schematic picture of the link integration in the LO is
shown in Fig.\ref{fig:link-integration}(a).
By the link integration,
on-site four-Fermi interaction term is generated,
which corresponds to the on-site repulsion between opposite spins (Hubbard term).
In order to control the strength of this interaction for later purpose,
we introduce an overall coefficient $z (>0)$.
Thus, the effective action can be written in terms of fermionic fields $a$ and $b$ as
\begin{eqnarray}
S_F^{(0)} \!\!\! &=& \!\! -\frac{z}{4}\left[ \sum_{\bfr \in A;\tau'} \!\!\!\! n_a(x) n_a(x+a\hat{\tau'}) + \!\!\!\! \sum_{\bfr \in B;\tau'} \!\!\!\! n_b(x) n_b(x+a\hat{\tau'}) \right] \nonumber \\
 && +\frac{2}{3} \sum_{\bfr \in A;\tau'} \sum_{i=1}^{3} \left[a^\dag(x) b(x+\bfs_i) + b^\dag(x+\bfs_i) a(x) \right] \!\!, \label{eq:sf0}
\end{eqnarray}
where $n_\chi(x)\equiv \chi^\dag(x)\chi(x)$ denotes the local charge density at the site $x=(\bfr,\tau')$.

Here we apply Stratonovich--Hubbard transformation
by introducing the bosonic auxiliary field $\sigma$,
which corresponds to the charge density difference between A and B sublattices, $\langle n_a-n_b \rangle$.
By mean-field approximation over $\sigma$,
the first line in Eq.(\ref{eq:sf0}) is converted into fermion bilinears as 
\begin{equation}
\frac{z}{2} \sum_{\bfr \in A \cup B;\tau'} \sigma^2 - \frac{z\sigma}{2} \left[ \sum_{\bfr \in A;\tau'} n_a(x) - \sum_{\bfr \in B;\tau'} n_b(x) \right]. \label{eq:eff-mass}
\end{equation}
Thus,
we can integrate out all the fermionic degrees of freedom,
to obtain the effective potential of this system at LO
per one pair of A and B sites:
\begin{eqnarray}
F_\mathrm{eff}^{(0)}(\sigma) &=& -\frac{1}{N_{\tau'} V} \ln Z^{(0)} \label{eq:feff0}\\
 &=& \frac{z}{2}\sigma^2 - \frac{1}{V}\int_{\bfk\in\Omega} d^2\bfk \ln\left[\left(\frac{z \sigma}{2}\right)^2 + \left|\frac{2}{3}\Phi(\bfk)\right|^2\right], \nonumber
\end{eqnarray}
where $V$ is the number of A (B) sites in the system.
$\int_{\bfk \in \Omega}$ is the integration over the Brillouin zone $\Omega$,
with normalization $\frac{1}{V} \int_{\bfk \in \Omega} d^2\bfk \cdot 1 =1$.
$a N_{\tau'}$ is the temporal lattice size, corresponding to the inverse temperature.
In this work we consider the zero-temperature and infinite volume limit, so that $N_{\tau'}$ and $V$ are set to infinity.
The first term in Eq.(\ref{eq:feff0}) represents the tree level of $\sigma$,
while the second logarithmic term comes from the one-loop effect of the fermion.

\subsection{Next-to leading order: $O(\beta^1)$}
Next, we consider the next-to LO (NLO) terms in the strong coupling expansion, $Z^{(1)}$.
At $O(\beta^1)$, one plaquette from $S_G$ contributes to the link integration.
$S_G$ (with instantaneous approximation) contains two kinds of plaquettes, $U_{i\tau'}$ and $U_{z\tau'}$,
but $U_{z\tau'}$ does not contribute to the link integration,
because the link in the $z$-direction cannot be canceled by the fermion hopping terms.
On the other hand, $U_{i\tau'}(x) \equiv U_{\tau'}(x+\bfs_i) U_{\tau'}^*(x)$ contributes to the link integration,
combined with two fermion hopping terms:
\begin{widetext}
\begin{eqnarray}
&& \int dU_{\tau'}(x) dU_{\tau'}(x+\bfs_i) e^{ -\frac{1}{2}\left[a^\dag(x)U_{\tau'}(x)a(x+a\hat{\tau'}) + b^\dag(x+\bfs_i)U_{\tau'}(x+\bfs_i)b(x+\bfs_i+a\hat{\tau'}) - \mathrm{H.c.}\right] } \times \frac{\sqrt{3}\beta}{2} U_{\tau'}(x+\bfs_i) U_{\tau'}^*(x) \nonumber \\
&=& \frac{\sqrt{3}\beta}{2} \int dU_{\tau'}(x) \left[ U_{\tau'}^*(x) - \frac{1}{2}a^\dag a' + \frac{1}{2}a'^\dag{U_{\tau'}^*}^2(x)a - \frac{1}{4}a^\dag a' a'^\dag U_{\tau'}^*(x)a \right] \nonumber \\
 && \qquad \times \int dU_{\tau'}(x+\bfs_i) \left[ U_{\tau'}(x+a_{\tau'}) - \frac{1}{2}b^\dag U_{\tau'}(x+a_{\tau'})^2 b' + \frac{1}{2}b'^\dag b - \frac{1}{4}b^\dag U_{\tau'}(x+a\hat{\tau'}) b' b'^\dag b \right] \label{eq:link-integration-1}\\
&=& -\frac{\sqrt{3}\beta}{8}a^\dag(x)a(x+a\hat{\tau'}) b^\dag(x+\bfs_i+a\hat{\tau'}) b(x+\bfs_i) ,
\end{eqnarray}
\end{widetext}
 as shown in Fig.\ref{fig:link-integration}(b).
(In Eq.(\ref{eq:link-integration-1}), we denote $a\equiv a(x),\; a'\equiv a(x+a\hat{\tau'}), \; b\equiv b(x+\bfs_i)$ and $b' \equiv b(x+\bfs_i+a\hat{\tau'})$.)
Thus, the effective action in the NLO can be written in terms of fermions as
\begin{eqnarray}
S_F^{(1)} \!\! &=& \!\! -\xi \!\!\! \sum_{\bfr\in A;\tau'} \sum_{i=1}^{3}\\
&&  \left[a^\dag(x) b(x+\bfs_i) b^\dag(x+\bfs_i+a\hat{\tau'}) a(x+a\hat{\tau'}) +\mathrm{H.c.} \right]. \nonumber
\end{eqnarray}
Hereafter, we use the rescaled parameter $\xi \equiv \sqrt{3}\beta/8$ instead of $\beta$
as the strength of such a nearest-neighbor interaction.

In order to convert this interaction term into fermion bilinears,
we apply the ``extended'' Stratonovich--Hubbard transformation,
\begin{equation}
\int e^{\alpha AB} = \mathrm{const.} \times \int d\lambda d\lambda^* e^{-\alpha[|\lambda|^2-\lambda A-\lambda^* B]},
\end{equation}
with the positive constant $\alpha$ and the complex auxiliary field $\lambda$.
With the auxiliary field $\lambda_i(\bfr,\tau')$
corresponding to the fermion bilinear $\langle a^\dag(x) b(x+\bfs_i)\rangle$,
we obtain the NLO effective action
\begin{equation}
S_F^{(1)}=
2\xi \!\!\! \sum_{\bfr\in A;\tau'} \sum_{i=1}^{3} \left[|\lambda_i(x)|^2 - \left( \lambda_i(x) a^\dag(x)b(x+\bfs_i) + \mathrm{H.c.} \right)\right].
\end{equation}
Thus, $S_F^{(1)}$ modifies the hopping of the fermions in the spatial direction
through the auxiliary field $\lambda_i$.

\begin{figure}[tb]
\begin{center}
\includegraphics[width=8cm]{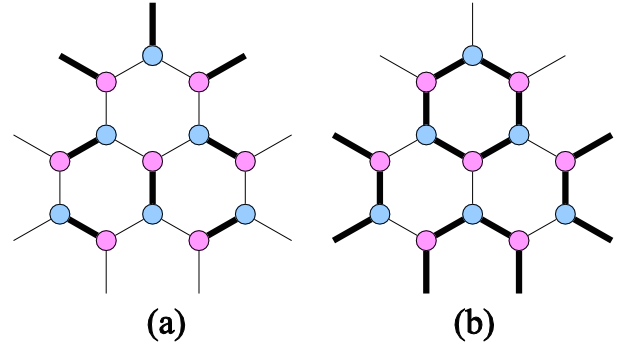}
\end{center}
\vspace{-0.5cm}
\caption{Schematic picture of the Kekul\'e distortion pattern.
Thick lines and thin lines represent the strong hopping and the weak hopping, respectively.
(a) Distortion pattern for $\lambda_\Delta>0$.
(b) Distortion pattern for $\lambda_\Delta<0$.}
\label{fig:kekulepattern}
\end{figure}

Here,
we take the ansatz that $\lambda_i$ should be split into the spatially uniform part and the spatially varying part with the Kekul\'e distortion pattern:
\begin{equation}
\lambda_j(\bfr,\tau') \equiv \lambda_\sigma + \lambda_\Delta e^{2\pi i/3} \left[e^{i(\bfK_+\cdot\bfs_j + \bfG\cdot\bfr)} + e^{i(\bfK_-\cdot\bfs_j - \bfG\cdot\bfr)}\right],
\end{equation}
where $\bfG \equiv \bfK_+ - \bfK_-$,
and $\lambda_\sigma$ and $\lambda_\Delta$ are real values.
The first term renormalizes the Fermi velocity $\vf$ uniformly,
with the factor $Z_v \equiv 1-\xi\lambda_\sigma/3$.
We show later that $\langle \lambda_\sigma \rangle <0$,
so that the Fermi velocity becomes faster at finite $\beta$ (or $\xi$)
than that in the strong coupling limit $(\beta=0)$.
The second term corresponds to the spontaneous Kekul\'e distortion,
with the amplitude $\Delta = 3\xi\lambda_\Delta$.
The Kekul\'e distortion is characterized by the pattern of alternating bond strengths,
as shown in Fig.\ref{fig:kekulepattern},
and induces a spectral gap without breaking the sublattice (chiral) symmetry \cite{Hou_2007},
with the modified dispersion relation $E(\bfK_\pm+\bfk)=\sqrt{|\bfk|^2+|\Delta|^2}+O(\bfk^4)$.
Since its unit lattice is three times as large as that of the ordinary honeycomb lattice in the real space,
the Brillouin zone $\Omega$ is split into three hexagonal parts:
$\tilde{\Omega}$ and $\tilde{\Omega}_\pm$,
surrounding $\Gamma$-point and the Dirac points $\bfK_\pm$ respectively,
as shown in Fig.\ref{fig:brillouinzone2}.
As a result,
the effective action up to the NLO can be written with the order parameters $\sigma$, $\lambda_\sigma$ and $\lambda_\Delta$ as
\begin{eqnarray}
&& S_F^{(0)}+S_F^{(1)} = \sum_{\bfr\in A;\tau'} \left[\frac{z}{2}\sigma^2 + 6\xi(\lambda_\sigma^2 +2\lambda_\Delta^2)\right] \\
&& + \sum_{\bfk \in \tilde{\Omega},\tau'} \tilde{\Psi}^\dag(\bfk,\tau') \left( \begin{array}{cc}
   -(z/2)\sigma I_3 & (2/3)\tilde{\Phi}^\dag(\bfk) \\
   (2/3)\tilde{\Phi}(\bfk) & (z/2)\sigma I_3
  \end{array}\right)
 \tilde{\Psi}(\bfk,\tau'), \nonumber
\end{eqnarray}
where the $3\times 3$ matrix $\tilde{\Phi}(\bfk) \equiv \tilde{\Phi}_0(\bfk) - 3\xi\tilde{\Phi}_1(\bfk)$, with
\begin{eqnarray}
\tilde{\Phi}_0(\bfk) &\equiv& \left( \begin{array}{ccc}
   \Phi(\bfk) & 0 & 0 \\
   0 & \Phi(\bfK_+ +\bfk) & 0 \\
   0 & 0 & \Phi(\bfK_- +\bfk)
  \end{array}\right),\\
\tilde{\Phi}_1(\bfk) &\equiv& \left( \begin{array}{ccc}
   \lambda_\sigma \Phi(\bfk) & \!\! \lambda'_\Delta \Phi(\bfK_- \!\!+\bfk) & \!\! \lambda'_\Delta \Phi(\bfK_+ \!\!+\bfk) \\
   \!\! \lambda'_\Delta \Phi(\bfK_- \!\!+\bfk) & \!\! \lambda_\sigma \Phi(\bfK_+ \!\!+\bfk) & \lambda'_\Delta \Phi(\bfk) \\
   \!\! \lambda'_\Delta \Phi(\bfK_+ \!\!+\bfk) & \lambda'_\Delta \Phi(\bfk) & \!\! \lambda_\sigma \Phi(\bfK_- \!\!+\bfk)
  \end{array}\right), \nonumber
\end{eqnarray}
and $I_3$ is a $3\times 3$ unit matrix.
Here we denote $\lambda'_\Delta \equiv \lambda_\Delta e^{-2\pi i/3}$ for simplicity.
The fermionic field $\tilde{\Psi}$ is defined by
$
\tilde{\Psi}(\bfk,\tau') \equiv \bigl[a(\bfk,\tau'),a(\bfK_+ +\bfk,\tau'),a(\bfK_- +\bfk,\tau') ,b(\bfk,\tau'),b(\bfK_+ +\bfk,\tau'),b(\bfK_- +\bfk,\tau')\bigr]^T.
$
By integrating out the fermion field $\tilde{\Psi}$,
we obtain the effective potential
\begin{eqnarray}
&& F_\mathrm{eff}^{(0+1)}(\sigma,\lambda_\sigma,\lambda_\Delta) = \frac{z}{2}\sigma^2 +6\xi(\lambda_\sigma^2+2\lambda_\Delta^2) \label{eq:feff1}\\
&& \quad - \frac{1}{V}\int_{\bfk\in\tilde{\Omega}} d^2\bfk \ln\det\left[\left(\frac{z\sigma}{2}\right)^2 I_3 + \left(\frac{2}{3}\right)^2 \tilde{\Phi}^\dag(\bfk)\tilde{\Phi}(\bfk)\right]. \nonumber
\end{eqnarray}
The third term (fermion loop effect) is modified from that in Eq.(\ref{eq:feff0})
by the spontaneous Kekul\'e distortion $\lambda_\Delta$.

\section{Qualitative properties}\label{sec:analysis}
So far we have reconstructed the effective fermion model
with two interaction terms,
the on-site interaction and the NN interaction,
obtained by the strong coupling expansion of the U(1) lattice model,
and derived the effective potential of the system:
$F_\mathrm{eff}^{(0+1)}(\sigma,\lambda_\sigma,\lambda_\Delta)$.
Hereafter,
we vary the strengths of these interaction terms ($z$ and $\xi$ respectively)
to arbitrary values,
to observe the interplay between the sublattice symmetry broken (SLSB) phase
and the Kekul\'e distortion (KD) phase.
First we investigate the qualitative properties of possible phases in the system
by taking the characteristic limits of $z$ and $\xi$:
the SLSB phase in the limit $\xi\sim 0$,
and the spontaneous KD phase in the limit $z=0$.
Then, we consider the competition between these two phases
by approximating the effective potential in the region
where both $z$ and $\xi$ are considerably small,
and estimate the phase structure of the system qualitatively.
As a result, we find that the appearance of the SLSB phase or the KD phase
is related to the dominance of the on-site term or the NN term respectively,
and that the KD phase is split into two phases (KD1 and KD2),
flipping the sign of $\lambda_\Delta$.

\subsection{Sublattice symmetry broken phase: $\xi\sim 0$}
First we consider the limit $\xi=0$,
where only the on-site interaction is concerned.
In this limit,
the effective potential in Eq.(\ref{eq:feff1}) reduces to the simpler one in Eq.(\ref{eq:feff0}).
The first term (tree level of $\sigma$) becomes dominant as $|\sigma|\rightarrow \infty$,
while the second term (fermion one-loop effect) dominates when $|\sigma| \rightarrow 0$.
Due to the logarithmic singularity of $\partial^2 F_\mathrm{eff}^{(0)}/\partial \sigma^2$
around $\sigma=0$ from the one-loop term,
$F_\mathrm{eff}^{(0)}(\sigma)$ has a minimum at finite $\sigma$ for any value of $z>0$.
The potential minimum gives the expectation value of the charge density imbalance between two sublattices,
$\langle \sigma \rangle = \langle n_a - n_b \rangle$,
which serves as the order parameter of the spontaneous sublattice (chiral) symmetry breaking.
In the 4-component Dirac fermion representation,
it corresponds to the chiral condensate $\langle \bar{\psi}\psi \rangle$.
Therefore,
the sublattice symmetry of the system is spontaneously broken in the limit $\xi=0$.

\begin{figure}[tb]
\begin{center}
\includegraphics[width=7.5cm]{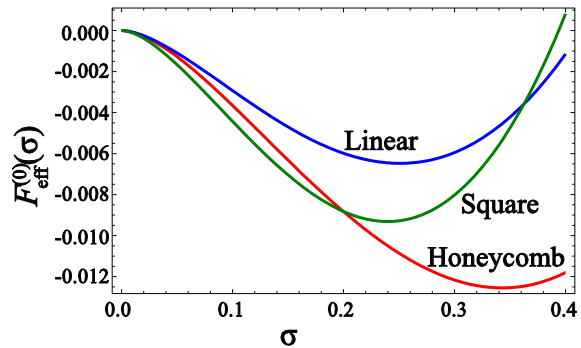}
\end{center}
\vspace{-0.5cm}
\caption{The behavior of the effective potential of the system in the strong coupling limit, $F_\mathrm{eff}^{(0)}$,
 as a function of exciton (chiral) condensate $\sigma$ at $z=1$.
Honeycomb: $F_\mathrm{eff}^{(0)}(\sigma)$ in Eq.(\ref{eq:feff0}) with the exact dispersion relation $\Phi(\bfk) = \sum_i e^{i\bfk\cdot\bfs_i}$.
Linear: $F_\mathrm{eff}^{(0)}(\sigma)$ in Eq.(\ref{eq:feff0}) with the approximate dispersion relation $\Phi(\bfK_\pm+\bfk) = \frac{3}{2}e^{-2\pi i/3} a (\pm k_x + ik_y)$.
Square: The effective potential obtained from the square lattice formulation [Eq.(\ref{eq:feff-sq})].}
\label{fig:feff_honeycomb1}
\end{figure}

The behavior of the effective potential in Eq.(\ref{eq:feff0}) at $z=1$ is shown in Fig.\ref{fig:feff_honeycomb1}
as the curve with the label ``Honeycomb.''
In Fig.\ref{fig:feff_honeycomb1},
$F_\mathrm{eff}^{(0)}(\sigma)$ obtained from two other formulations are displayed:
``Linear'' is the effective potential calculated with the Dirac cone approximation $\Phi(\bfK_\pm +\bfk)=\frac{3}{2}e^{-2\pi i/3}a(\pm k_x+ik_y)$,
and ``Square'' is the one obtained from the square lattice formulation \cite{Araki_2010},
\begin{equation}
F_\mathrm{eff}^{(0)}(\sigma) = \frac{\sigma^2}{2} -\frac{1}{(2\pi)^2} \int_{\bfk \in [-\pi,\pi]^2} \!\!\!\!\! \!\!\!\!\! \!\!\!\!\! \!\!\! d\bfk^2 \ln \left[\left(\frac{\sigma}{2}\right)^2 + \sum_{j=x,y} \sin^2 k_j \right]. \label{eq:feff-sq}
\end{equation}
All of them qualitatively have the same structure
because they have the Dirac cone structure in common around the Dirac points,
but quantitative behaviors are different due to the deviation from the Dirac cone structure
at large momentum.

As seen from Eq.(\ref{eq:eff-mass}),
finite $\sigma$ induces an on-site energy difference between two sublattice
in the sense of mean-field,
yielding a finite spectral gap $E(\bfK_\pm) =\vf z \sigma/2a$,
which corresponds to the dynamically generated mass of the fermion.
When $z$ takes the physical value $z=1$,
the expectation value is $\sigma=0.343$,
which gives the dynamical gap $0.72\mathrm{eV}$.
By taking the momentum integration around $\bfk=0$,
we have an approximate relation
\begin{equation}
\frac{\partial F_\mathrm{eff}^{(0)}}{\partial \sigma} \sim z+4\pi\sigma\left(\frac{z}{2}\right)^2 \ln\left(\frac{z\sigma}{2}\right)^2
\end{equation}
for sufficiently small $z$ and $\sigma$.
Thus, $z$-dependence of $\langle \sigma \rangle$ around $z$ can be approximated as
\begin{equation}
\langle \sigma \rangle \sim (2/z) \exp(-2/z),
\end{equation}
so that $\langle \sigma \rangle$ reaches toward $0$ as $z\rightarrow 0$.

Next,
we observe the behavior of the effective potential in the vicinity of $\xi=0$.
In order to take the effective potential up to $O(\xi^1)$,
we have to expand the fermion determinant by $\xi$.
By using the formula
\begin{equation}
\ln \det [A+\xi B] = \ln\det A + \Tr [\xi A^{-1}B] + O(\xi^2),
\end{equation}
the third term in Eq.(\ref{eq:feff1}) is approximated up to $O(\xi^1)$ as
\begin{eqnarray}
&& \ln\det\left[\left(\frac{z\sigma}{2}\right)^{2} +\left(\frac{2}{3}\right)^{2}(\tilde{\Phi}_0^\dag-3\xi\tilde{\Phi}_1^\dag)(\tilde{\Phi}_0-3\xi\tilde{\Phi}_1)\right] \nonumber \\
&\simeq& \ln\det\left[\left(\frac{z\sigma}{2}\right)^{2} +\left(\frac{2}{3}\right)^{2} \tilde{\Phi}_0^\dag \tilde{\Phi}_0 \right] \label{eq:expand-phi}\\
&& -3\xi\Tr\left[\left(\frac{z\sigma}{2}\right)^{\!\!2} +\left(\frac{2}{3}\right)^{\!\!2} \tilde{\Phi}_0^\dag \tilde{\Phi}_0\right]^{\! -1} \!\!\!\! \left(\frac{2}{3}\right)^{\!\!2} \! \left[\tilde{\Phi}_1^\dag \tilde{\Phi}_0 + \tilde{\Phi}_0^\dag \tilde{\Phi}_1\right]. \nonumber
\end{eqnarray}
Since the matrices $\left[(z\sigma/2)^2+(2/3)^2\tilde{\Phi}_0^\dag \tilde{\Phi}_0\right]^{-1}$ and $\tilde{\Phi}_0$ are diagonal,
only the diagonal part of $\tilde{\Phi}_1$,
which can be written as $\lambda_\sigma \tilde{\Phi}_0$,
contributes to the trace in the second term in Eq.(\ref{eq:expand-phi}).
Thus, we have the $\xi$-expansion of the effective potential as
$
F_\mathrm{eff}^{(0+1)}(\sigma,\lambda_\sigma,\lambda_\Delta) = F_\mathrm{eff}^{(0)}(\sigma) + F_\mathrm{eff}^{(1)}(\sigma,\lambda_\sigma,\lambda_\Delta) +O(\xi^2)
$,
where
\begin{eqnarray}
&& F_\mathrm{eff}^{(1)}(\sigma,\lambda_\sigma,\lambda_\Delta) = 6\xi(\lambda_\sigma^2+2\lambda_\Delta^2) \\
&& \quad \quad \quad +\frac{6\xi\lambda_\sigma}{V}\int_{\bfk\in \Omega} d^2\bfk \frac{(2/3)^2|\Phi(\bfk)|^2}{(z\sigma/2)^2+(2/3)^2|\Phi(\bfk)|^2}. \nonumber
\end{eqnarray}

Taking the potential minimum by the NN-auxiliary fields $\lambda_\sigma$ and $\lambda_\Delta$,
we obtain their expectation values up to the LO,
\begin{eqnarray}
\lambda_\Delta &=& 0 + O(\xi^1), \\
\lambda_\sigma &=& -\frac{1}{2V} \int_{\bfk\in \Omega} d^2\bfk \frac{(2/3)^2|\Phi(\bfk)|^2}{(z\sigma/2)^2+(2/3)^2|\Phi(\bfk)|^2} + O(\xi^1). \nonumber
\end{eqnarray}
Since $\lambda_\Delta$ does not contribute to the fermion one-loop term up to $O(\xi^1)$,
the Kekul\'e distortion does not appear around the limit $\xi=0$.
On the other hand,
$\lambda_\sigma$ acquires a negative expectation value,
so that the renormalization factor of the Fermi velocity, $Z_v=1-3\xi\lambda_\sigma$,
becomes larger than unity.
By substituting these relation to $F_\mathrm{eff}^{(1)}(\sigma,\lambda_\sigma,\lambda_\Delta)$,
the NLO effective potential can be rewritten as a function only of $\sigma$:
\begin{equation}
F_\mathrm{eff}^{(1)}(\sigma) = -\frac{3}{2}\xi \left[\frac{1}{V} \int_{\bfk\in \Omega} d^2\bfk \frac{(2/3)^2|\Phi(\bfk)|^2}{(z\sigma/2)^2+(2/3)^2|\Phi(\bfk)|^2}\right]^2.
\end{equation}
Since this term monotonically increases as a function of $\sigma$,
it reduces the expectation value of $\sigma$
(that is, the position of the potential minimum).
At the physical value $z=1$,
the expectation value of $\sigma$ is given up to NLO as
\begin{equation}
\sigma(\xi) = 0.342-1.73\xi +O(\xi^2),
\end{equation}
and $\lambda_\sigma = -0.471 +O(\xi)$.
Therefore,
the system reveals the SLSB phase in the vicinity of $\xi=0$,
and the amplitude of SLSB (charge density imbalance between two sublattices), $\sigma$,
decreases as a function of $\xi$.

\subsection{Kekul\'e distortion phase: $z=0$}
In order to investigate the qualitative properties of the Kekul\'e distortion (KD) phase,
we take the limit $z=0$,
where the system does not contain the on-site interaction
so that it may not reveal the SLSB phase.
Here, the effective potential reads
\begin{eqnarray}
&& F_\mathrm{eff}^{(0+1)}(\lambda_\sigma,\lambda_\Delta) \nonumber \\
&& = 6\xi(\lambda_\sigma^2+2\lambda_\Delta^2) - \frac{2}{V}\int_{\bfk\in\tilde{\Omega}} d^2\bfk \ln\det\left| \frac{2}{3} \tilde{\Phi}(\bfk)\right| \\
&& = 6\xi(\lambda_\sigma^2+2\lambda_\Delta^2) - \frac{2}{V}\int_{\bfk\in\tilde{\Omega}} d^2\bfk\cdot \label{eq:feff01-z}\\
&& \ln \Biggl|\left[\left(\frac{2}{3}Z_v\right)^3-2(2\xi\lambda_\Delta)^3\right]\Phi(\bfk)\Phi(\bfK_+ +\bfk)\Phi(\bfK_- +\bfk) \nonumber \\
&& -\frac{2}{3}Z_v (2\xi\lambda_\Delta)^2 e^{2\pi i/3}\left[\Phi^3(\bfk)+\Phi^3(\bfK_+ +\bfk)+\Phi^3(\bfK_- +\bfk)\right] \Biggr| \nonumber
\end{eqnarray}
First we analyze whether $\lambda_\Delta$ takes a finite expectation value or not.
Since one can easily see $\partial F_\mathrm{eff}^{(0+1)}/\partial \lambda_\Delta |_{\lambda_\Delta=0} =0$,
$\lambda_\Delta=0$ is either a local maximum or minimum of the effective potential.
In order to consider the behavior around $\lambda_\Delta$,
we have to check the sign of the second derivative
\begin{eqnarray}
&& \frac{\partial^2 F_\mathrm{eff}^{(0+1)}}{\partial \lambda_\Delta^2} \Biggr|_{\lambda_\Delta=0} = 24\xi +\frac{2}{V}\int_{\bfk \in \tilde{\Omega}}d^2\bfk\cdot \\
&& \quad \quad \quad \frac{\frac{16}{3}Z_v \xi^2 [\Phi^3(\bfk) +\Phi^3(\bfK_+ +\bfk) +\Phi^3(\bfK_- +\bfk)]}{(\frac{2}{3}Z_v)^3 e^{-2\pi i/3}\Phi(\bfk)\Phi(\bfK_+ +\bfk)\Phi(\bfK_- +\bfk)} . \nonumber
\end{eqnarray}
Since the denominator of the integrand becomes zero only at $\bfk=0$,
the region around this point is dominant in the loop integration.
Taking the leading order in $|\bfk|$ in the numerator and the denominator,
the loop integral becomes
\begin{equation}
-\frac{2}{V}\int_{\bfk \in \tilde{\Omega}} d^2\bfk \frac{144 Z_v \xi^2 +O(|\bfk|^2)}{2 Z_v |\bfk|^2 +O(|\bfk|^3)},
\end{equation}
which has a negative logarithmic divergence.
Due to this logarithmic divergence in the momentum integration,
the sign of the second derivative becomes negative at $\lambda_\Delta=0$,
so that $\lambda_\Delta=0$ is a local maximum of the effective potential.
Therefore, for any value of $\xi(>0)$ (or $\beta$),
$\lambda_\Delta$ takes a finite expectation value.


Next, we consider the behavior of the potential minimum $(\lambda_\sigma,\lambda_\Delta)$.
Due to the logarithmic divergence of the momentum integration
when the order parameters satisfy the relation
$\frac{2}{3}-2\xi\lambda_\sigma - 2\sqrt[3]{2}\xi\lambda_\Delta =0$ [see Eq.(\ref{eq:feff01-z})],
the effective potential has a non-analyticity on this curve,
separating the $(\lambda_\sigma,\lambda_\Delta)$-plane into two regions.
In each region there is a local minimum of $F_\mathrm{eff}$,
and it depends on the value of $\xi$ which minimum is taken.
When $\xi$ crosses over a certain value $\xi_K$,
one local potential minimum may dominate over the other one,
causing a sudden jump of $\langle \lambda_\Delta \rangle$.
Therefore,
the KD phase is split into two regions at the line $\xi=\xi_K$,
where the system reveals the first order phase transition.
Here we refer to these two  phases
as KD1 for $\xi<\xi_K$ and KD2 for $\xi>\xi_K$, respectively.

Finally,
we consider the behavior of $\langle \lambda_\Delta \rangle$ in the limits
$\xi\sim 0$ and $\xi \rightarrow \infty$.
In the limit $\xi\sim 0$,
we neglect the $\lambda_\sigma$-dependence
because it depends on the loop integration only via $Z_v=1-3\xi\lambda_\sigma$,
which becomes unity at $\xi=0$.
By performing the momentum integration around $\bfk=0$,
we have the approximate relation
\begin{equation}
F_\mathrm{eff}^{(0+1)}(\lambda_\Delta) \sim 12\xi\lambda_\Delta^2 + A(2\xi\lambda_\Delta)^2 \ln (2\xi\lambda_\Delta)^2,
\end{equation}
where $A$ is a positive constant related to the area of the momentum integration.
By taking the potential minimum,
we can estimate the order of $\langle \lambda_\Delta \rangle$ to be
\begin{equation}
\langle \lambda_\Delta \rangle \sim \xi^{-1} \exp (-\xi^{-1}).
\end{equation}
Therefore, $\langle \lambda_\Delta \rangle$ reaches toward zero as $\xi \rightarrow 0$.
On the other hand, in the limit $\xi\rightarrow\infty$,
all the terms in $|\cdots|$ in Eq.(\ref{eq:feff01-z}) becomes proportional to $\xi^3$.
Thus, the $\xi$-dependence in the logarithm can be factored out,
so that only the first term (tree level of $\lambda_\sigma$ and $\lambda_\Delta$) becomes dominant in this limit.
Therefore,
both the expectation values of $\lambda_\Delta$ and $\lambda_\sigma$ reach toward zero in the limit $\xi\rightarrow\infty$.

\subsection{Competition between SLSB and KD phases}
Let us now investigate the competition between two phases, SLSB and KD,
and observe what kind of phase transition may occur between these two phases.
In order to treat the logarithmic singularity of the loop integral,
we take into account the momentum space only around the Dirac points.
Here we consider the region
where the interaction strengths $z$ and $\xi$ are sufficiently small,
to simplify the discussion.
Since $\xi\lambda_\Delta$ reaches toward zero as $\xi\rightarrow 0$,
we assume that the terms of $O(\xi\lambda_\Delta|\bfk|)$ and the smaller ones are negligible,
which gives a simplified form as follows:
\begin{eqnarray}
&& \left(\frac{z\sigma}{2}\right)^2 I_3 +\left(\frac{2}{3}\right)^2\tilde{\Phi}^\dag(\bfk) \tilde{\Phi}(\bfk) \nonumber \\
&& \simeq \mathrm{diag}\Bigl\{ \left(\frac{z\sigma}{2}\right)^2+4Z_v^2, \left(\frac{z\sigma}{2}\right)^2+|Z_v \bfk|^2+36(\xi\lambda_\Delta)^2, \nonumber \\
&& \quad \quad \quad \quad \quad \left(\frac{z\sigma}{2}\right)^2+|Z_v \bfk|^2+36(\xi\lambda_\Delta)^2 \Bigr\}.
\end{eqnarray}
This simplification is valid as long as the logarithmic singularity is dominant,
that is, $\sigma$ and $\xi\lambda_\Delta$ are in vicinity of $0$.
Since the first element of this matrix does not contribute to the logarithmic singularity,
we can further simplify this model by neglecting the first element
(contribution from the Brillouin zone $\tilde{\Omega}$, which does not cover the Dirac points $\bfK_\pm$).
Thus, we obtain the effective potential
\begin{eqnarray}
&& F_\mathrm{eff}^{(0+1)}(\sigma,\lambda_\sigma,\lambda_\Delta) \simeq \frac{z}{2}\sigma^2 +6\xi(\lambda_\sigma^2+2\lambda_\Delta^2) \label{eq:feff01-simple}\\
&& \quad \quad - \frac{2}{V}\int_{\bfk\in\tilde{\Omega}} d^2\bfk \ln\left[\left(\frac{z\sigma}{2}\right)^2+36\xi^2\lambda_\Delta^2+|Z_v\bfk|^2\right]. \nonumber
\end{eqnarray}

The properties of the effective potential in Eq.(\ref{eq:feff01-simple}) can be observed rather easily than the exact one.
If we define a new field $\phi$ by
$
\phi^2 \equiv \sigma^2 + (144\xi^2/z^2) \lambda_\Delta^2,
$
the effective potential is rewritten as
\begin{eqnarray}
F_\mathrm{eff}(\phi,\lambda_\sigma,\lambda_\Delta) &=& \frac{z}{2}\phi^2 + 12\xi\left(1-\frac{6\xi}{z}\right)\lambda_\Delta^2 + 6\xi\lambda_\sigma^2 \label{eq:feff-phi} \\
&& - \frac{2}{V}\int_{\bfk\in\tilde{\Omega}} d^2\bfk \ln\left[\left(\frac{z\phi}{2}\right)^2 +|Z_v \bfk|^2\right]. \nonumber
\end{eqnarray}
If $1-6\xi/z=0 \; (z=6\xi)$,
the effective potential is given as a function of $\phi$ and $\lambda_\sigma$,
and does not depend on $\lambda_\Delta$ explicitly.
Therefore, $\sigma$ and $\lambda_\Delta$ can take arbitrary expectation values satisfying
\begin{equation}
\langle \sigma \rangle^2 + \frac{144\xi^2}{z^2}\langle \lambda_\Delta \rangle^2 = \langle \phi \rangle^2.
\end{equation}
If $1-6\xi/z>0 \; (z>6\xi)$,
the second term in Eq.(\ref{eq:feff-phi}) behaves as a symmetry breaking term between $\sigma$ and $\lambda_\Delta$.
Since the fermion loop integral does not explicitly depend on $\lambda_\Delta$,
the effective potential monotonically increases as a function of $\lambda_\Delta$,
yielding $\langle \lambda_\Delta \rangle =0$.
On the other hand, $\phi$ contributes to the loop integral, so that $\langle \phi \rangle \neq 0$.
Therefore, we have $\langle \sigma \rangle \neq 0$,
that is, the sublattice symmetry is spontaneously broken.

If $1-6\xi/z<0 \; (z<6\xi)$,
the coefficient of the second term in Eq.(\ref{eq:feff-phi}) becomes negative,
leading to the unphysical result $\langle \lambda_\Delta \rangle=\infty$.
Here we rewrite the effective potential as a function of $\phi$,$\sigma$ and $\phi_\sigma$,
so that all the coefficients of these variables would be positive:
\begin{eqnarray}
F_\mathrm{eff}(\phi,\sigma,\lambda_\sigma) &=& \frac{z^2}{12\xi}\phi^2 + \frac{z}{2} \left(1-\frac{z}{6\xi}\right)\sigma^2 + 6\xi\lambda_\sigma^2 \label{eq:feff-phi'} \\
&& - \frac{2}{V}\int_{\bfk\in\tilde{\Omega}} d^2\bfk \ln\left[\left(\frac{z\phi}{2}\right)^2 +|Z_v \bfk|^2\right]. \nonumber
\end{eqnarray}
In this form,
the effective potential monotonically rises as a function of $\sigma$,
so that we have $\langle \sigma \rangle=0$, $\langle \phi \rangle \neq 0$ and $\langle \lambda_\Delta \rangle \neq 0$.
In other words,
there appears a Kekul\'e distortion pattern spontaneously.

\begin{figure}[tb]
\begin{center}
\includegraphics[width=7cm]{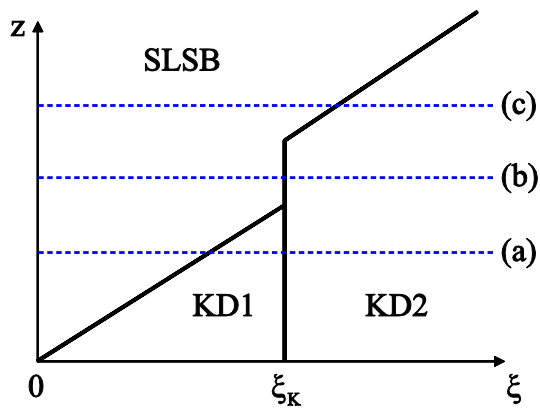}
\end{center}
\vspace{-0.5cm}
\caption{A hypothetical phase diagram of the effective model of monolayer graphene,
from the qualitative investigation of the effective potential $F_\mathrm{eff}^{(0+1)}$.
When the on-site interaction ($z$) is dominant over the nearest-neighbor interaction ($\xi$),
the system spontaneously shows the sublattice symmetry broken (SLSB) phase;
otherwise the system shows the Kekul\'e distortion (KD) phase.
The KD phase is separated by the line $\xi=\xi_K$ into two regions,
which correspond to the two local minima of the effective potential respectively.
All the critical line show the first-order transition behavior.
The dashed lines (a), (b) and (c) correspond to the results from the numerical calculation:
(a) corresponds to $z=1$ in Fig.\ref{fig:sigma-beta-z1},
(b) to $z=40$ in Fig.\ref{fig:sigma-beta-z40},
and (c) to $z=50$ in Fig.\ref{fig:sigma-beta-z15}.}
\label{fig:phasediagram}
\end{figure}

Therefore,
when crossing the line $z=6\xi$,
there is a first-order phase transition between the sublattice (chiral) symmetry broken (SLSB) phase
and the spontaneous Kekul\'e distortion (KD) phase.
Moreover, as shown in the previous subsection,
the expectation value of $\lambda_\Delta$ reveals the non-analyticity at a certain value $\xi_K$ in the KD phase.
Since the effective potential does not depend on $z$ in the KD phase,
the value of $\xi_K$ is independent of $z$ as long as the point $(z,\xi_K)$ is in the KD phase.
Since the KD1 and the KD2 phases correspond to different potential minima respectively,
the critical line between SLSB and KD1 and that between SLSB and KD2 are discontinuous.
From the qualitative discussions above,
we can map a schematic phase diagram of the system, as shown in Fig.\ref{fig:phasediagram}.

\section{Numerical results}\label{sec:numerical}
\begin{figure}[tb]
\begin{center}
\includegraphics[width=8.5cm]{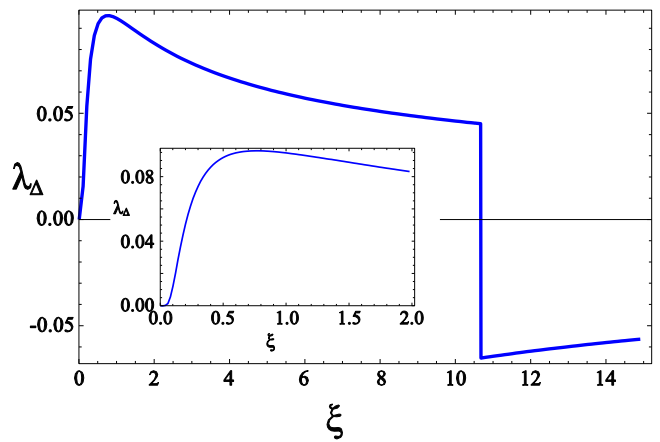}
\end{center}
\vspace{-0.5cm}
\caption{The behavior of the spontaneous Kekul\'e distortion (KD) $\lambda_\Delta$
as a function of the nearest-neighbor interaction strength $\xi$,
in the absence of the on-site interaction ($z=0$).
The inset shows the behavior of $\lambda_\Delta$ in vicinity of $\xi=0$.
$\lambda_\Delta$ shows a non-analyticity at $\xi_K=10.67$,
where the KD phase is separated into two regions, KD1 ($\lambda_\Delta>0$) and KD2 ($\lambda_\Delta<0$).
}
\label{fig:z0}
\end{figure}

Now we confirm the qualitative results above
by minimizing the exact effective potential in Eq.(\ref{eq:feff1}) numerically.
In the limit $z=0$, the effective potential becomes independent of $\sigma$,
so that we only derive the expectation value of $\lambda_\Delta$ as a function of $\xi$,
as shown in Fig.\ref{fig:z0}.
It can be clearly seen that $\lambda_\Delta(\xi)$ shows a non-analyticity at $\xi_K=9.97$,
which separates the KD phase into two regions.
In the KD1 region ($\xi<\xi_K$),
$\lambda_\Delta$ obtains a positive expectation value,
which corresponds to the lattice distortion pattern shown in Fig.\ref{fig:kekulepattern}(a).
As qualitatively estimated,
$\langle \lambda_\Delta \rangle$ starts from zero at $\xi=0$
and monotonically increases for small $\xi$.
It has a peak at $\xi\sim 0.7$ and eventually decreases until $\xi=\xi_K$.
On the other hand, in the KD2 region ($\xi>\xi_K$),
$\lambda_\Delta$ takes a negative expectation value,
corresponding to the pattern in Fig.\ref{fig:kekulepattern}(b).
It then reaches toward zero as a function of $\xi$.
which agrees with the analytical result that $\langle \lambda_\Delta \rangle \rightarrow 0$ as $\xi\rightarrow\infty.$

\begin{figure}[tb]
\begin{center}
\includegraphics[width=8cm]{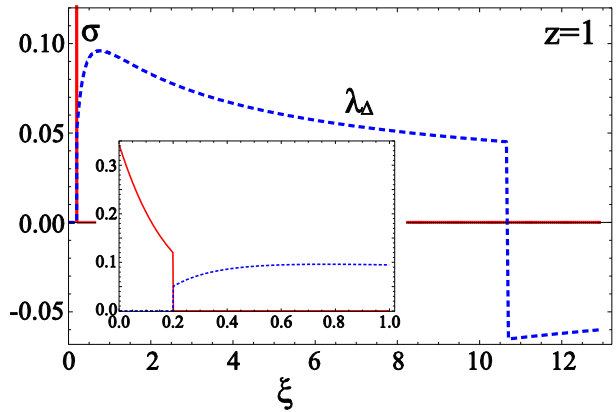}
\end{center}
\vspace{-0.5cm}
\caption{Expectation values of the spontaneous sublattice (chiral) symmetry breaking $\sigma$
and the Kekul\'e distortion amplitude $\lambda_\Delta$, calculated at the physical value $z=1$.
The inset shows the behavior of $\sigma$ and $\lambda_\Delta$ in vicinity of $\xi=0$.
There is a first order phase transition from the SLSB phase into the KD1 phase at $\xi=0.20$,
and that from KD1 into KD2 at $\xi_K=10.67$.
Such a behavior corresponds to the line (a) in the phase diagram in Fig.\ref{fig:phasediagram}.}
\label{fig:sigma-beta-z1}
\end{figure}

Next, we fix the on-site interaction strength $z$ to finite values.
At the value $z=1$,
which corresponds to the strong coupling expansion of the gauged model,
the expectation values of the sublattice symmetry breaking amplitude $\sigma$
and the spontaneous Kekul\'e distortion $\lambda_\Delta$ vary as function of $\xi$,
as shown in Fig.\ref{fig:sigma-beta-z1}.
The order parameters reveal non-analyticity at two points,
$\xi=0.20$ and $\xi=\xi_K$.
In the region $\xi < 0.20$,
$\sigma$ is finite and monotonically decreases,
while $\lambda_\Delta$ is zero.
As can be clearly seen from the analytic observation,
this region corresponds to the sublattice symmetry broken (SLSB) phase
in the hypothetical phase diagram in Fig.\ref{fig:phasediagram}.
In the region $0.20 < \xi < \xi_K$,
the expectation value of $\sigma$ vanishes,
while $\lambda_\Delta$ acquires a positive expectation value,
which corresponds to the KD1 phase.
For large value of $\xi$,
$\lambda_\Delta$ shows a non-analyticity at $\xi=\xi_K$,
obtains a negative expectation value, and reaches toward zero
just as seen in the $z=0$ limit.
This region corresponds to the KD2 phase.
Therefore, we can conclude that the axis $z=1$ corresponds to line (a)
in Fig.\ref{fig:phasediagram},
that is, the system turns from SLSB into KD1 at a certain critical value of $\xi$ (here $\xi=0.20$)
and turns from KD1 into KD2 at $\xi=\xi_K$.

\begin{figure}[tb]
\begin{center}
\includegraphics[width=8cm]{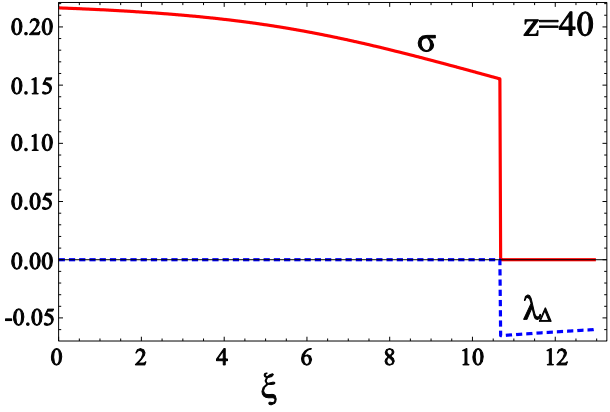}
\end{center}
\vspace{-0.5cm}
\caption{Expectation values of the spontaneous sublattice (chiral) symmetry breaking $\sigma$
and the Kekul\'e distortion amplitude $\lambda_\Delta$, calculated at $z=40$.
There is a first order phase transition from the SLSB phase into the KD2 phase at $\xi_K=10.67$,
and the KD1 phase does not appear.
Such a behavior corresponds to the line (b) in the phase diagram in Fig.\ref{fig:phasediagram}.}
\label{fig:sigma-beta-z40}
\end{figure}

When $z=40$,
there appears only one phase boundary,
as shown in Fig.\ref{fig:sigma-beta-z40}.
The phase transition occurs at $\xi=\xi_K$,
from the SLSB phase $(\sigma\neq 0)$ into KD2 phase $(\lambda_\Delta<0)$,
and the KD1 phase does not appear.
This behavior corresponds to the line (b) in Fig.\ref{fig:phasediagram}.

\begin{figure}[tb]
\begin{center}
\includegraphics[width=8cm]{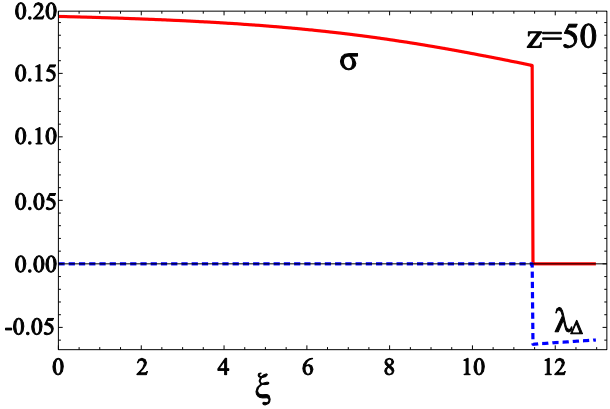}
\end{center}
\vspace{-0.5cm}
\caption{Expectation values of the spontaneous sublattice (chiral) symmetry breaking $\sigma$
and the Kekul\'e distortion amplitude $\lambda_\Delta$, calculated at $z=50$.
There is a first order phase transition from the SLSB phase into the KD2 phase at $\xi=11.45 (\neq \xi_K)$,
and the KD1 phase does not appear.
Such a behavior corresponds to the line (c) in the phase diagram in Fig.\ref{fig:phasediagram}.}
\label{fig:sigma-beta-z15}
\end{figure}

At the value $z=50$,
there appears only two phases as observed at $z=40$,
but here the critical value of $\xi$ is shifted from $\xi_K=10.67$,
as shown in Fig.\ref{fig:sigma-beta-z15}.
This behavior corresponds to the axis (c) in Fig.\ref{fig:phasediagram}.

\begin{figure}[tb]
\begin{center}
\includegraphics[width=8cm]{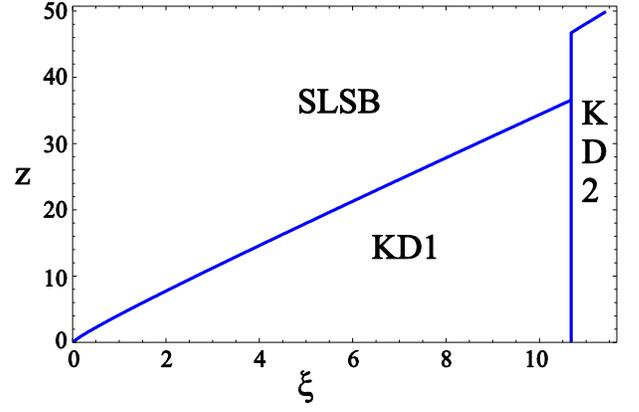}
\end{center}
\vspace{-0.5cm}
\caption{The phase boundary between the sublattice symmetry broken (SLSB) phase and the Kekul\'e distortion (KD) phase,
as a result of the numerical calculation.
As can be seen from Fig.\ref{fig:z0}, the KD phase is split into two phases by the line $\xi=10.67 (\equiv \xi_K)$:
KD1 $(\lambda_\Delta>0)$ and KD2 $(\lambda_\Delta<0)$.
This phase diagram agrees with the qualitative estimation obtained in Fig.\ref{fig:phasediagram}.}
\label{fig:xi-z}
\end{figure}

Finally, we show in Fig.\ref{fig:xi-z} the phase boundary between the SLSB and the KD phases,
which agrees with the qualitative phase diagram obtained in Fig.\ref{fig:phasediagram}.
In general,
the system turns into the KD phase when the nearest neighbor interaction $(\xi)$ becomes dominant
over the on-site interaction $(z)$,
that is, the Coulomb interaction strength becomes smaller.
Since the effective potential becomes independent of $z$ in the KD region,
there is a first order phase transition between KD1 and KD2 at the line $\xi=\xi_K$,
as seen in the $z=0$ limit.

\section{Conclusions and Outlook}\label{sec:conclusion}
In this work,
we have investigated the possible phase structure of monolayer graphene
with the on-site and the nearest neighbor (NN) interactions between fermions.
First,
the effective action of the system is constructed
including the electromagnetic field as U(1) link variables,
and the interaction terms between fermions are derived
by applying the techniques of the strong coupling expansion of the lattice gauge theory.
Thus we have obtained two kinds of effective interaction terms:
the on-site interaction which may contribute to the sublattice symmetry breaking (SLSB),
and the NN interaction which may lead to the Kekul\'e distortion (KD).
Using these two interaction terms,
we have reconstructed an effective model of graphene
with arbitrary interaction strengths $z$ and $\xi$ respectively,
to investigate the interplay between the SLSB and the KD.
We have observed the behavior of the order parameters with this effective model,
by the mean field approximation over the effective potential.

Focusing on the logarithmic singularity of the effective potential,
we have qualitatively obtained the phase diagram shown in Fig.\ref{fig:phasediagram}.
When the on-site interaction is dominant,
the sublattice (chiral) symmetry of the system is spontaneously broken,
leading to the dynamical mass term of the fermions.
On the other hand, when the nearest-neighbor interaction is sufficiently large,
the hopping parameters in the lattice get renormalized with the Kekul\'e distortion pattern.
In this case the fermions still obtain a dynamical spectral gap,
without breaking the sublattice (chiral) symmetry.
Moreover, this KD phase is split into two regions KD1 $(\lambda_\Delta>0)$ and KD2 $(\lambda_\Delta<0)$,
corresponding to two different minima of the effective potential.
Such a splitting line $\xi=\xi_K$ is numerically seen
by taking the limit where the on-site interaction is omitted ($z=0$).
For instance, when the on-site interaction strength $z=1$,
which corresponds to the strong coupling expansion of the Coulomb interaction,
the system reveals the SLSB phase in the strong coupling limit.
The system turns into the KD1 phase at $\xi=0.20$ ($\beta=0.92$),
and into the KD2 phase at $\xi=\xi_K=10.67$.
Since $z=1$ and $\beta=0.037 \; (\xi=0.008)$ in the vacuum-suspended monolayer graphene, 
we expect a gapped phase with SLSB,
while the system may reveal the KD phases if the Coulomb interaction is suppressed ($\beta$ is increased)
by the screening effect by substrates or the renormalization of the Fermi velocity $\vf$ \cite{Elias_2011}.
The KD1 phase does not appear at sufficiently large $z$,
as seen at $z=40$ and $50$ in this work.
It has been verified both qualitatively and numerically that
all the phase transitions in the phase diagram obtained in this work are first order phase transitions,
that is, the order parameters reveal non-analyticity when crossing the phase boundaries.

There are still several open questions to be solved within the framework of this study.
Since the spin degrees of freedom are absorbed in the fermion doubling,
which is the artifact of the lattice discretization,
spin-related ordering, such as the spin density wave (SDW) phase \cite{Semenoff_2011,Soriano_2011} and the ``spin-Kekul\'e'' phase \cite{Weeks_2010},
cannot be identified out of the SLSB and KD phases in this work.
Some other lattice discretization scheme that exactly treats the spin degrees of freedom is needed.
The effect beyond the NLO is also an interesting issue.
The next-to-NLO [$O(\beta^2)$] term,
which includes the four-Fermi interaction between second nearest neighboring sites,
can spontaneously generate an effective magnetic flux in the honeycomb plaquette,
leading to the so-called ``quantum anomalous Hall (QAH)'' state \cite{Raghu_2008}.
For example, Ref.\onlinecite{Giuliani_2012} treats several types of instabilities 
by the exact renormalization group method on the honeycomb lattice,
and shows that only four instabilities, SLSB, KD, SDW and QAH,
may occur by the effect of the Coulomb interaction,
but the competition among these orders is left for further investigation.
Extension of the lattice strong coupling expansion method to the bilayer graphene system is also required,
since a gapped phase has recently been observed in bilayer graphene experimentally \cite{ThomasWeitz_2010,Novoselov_2011}.
Quite a rich phase diagram is expected both in monolayer and bilayer graphene systems.

\
\begin{acknowledgments}
The author thanks H.~Aoki, C.~DeTar, T.~Hatsuda, K.~Nomura and S.~Sasaki for valuable comments and discussions.
This work is supported by Grant-in-Aid for Japan Society for the Promotion of Science (DC1, No.22.8037).

\end{acknowledgments}


\end{document}